\documentclass[10pt,preprint2]{aastex}
\usepackage{times}
\usepackage{epsfig}

\def \etal   {\hbox{et~al.\/}}

\def\lesssim{\mathrel{\hbox{\rlap{\hbox{\lower4pt\hbox{$\sim$}}}\hbox{$<$}}}}
\def\gtrsim{\mathrel{\hbox{\rlap{\hbox{\lower4pt\hbox{$\sim$}}}\hbox{$>$}}}}
\def\arcsec{\hbox{$^{\prime\prime}$}}

\begin{document}

\shorttitle{SPECTROPOLARIMETRY OF HD~92207}

\title{
SPECTROPOLARIMETRIC VARIABILITY AND CO-ROTATING STRUCTURE IN HD~92207}

\author{\sc R.~Ignace,}
\affil{
Department of Physics \& Astronomy,
East Tennessee State University, Johnson City, TN, 37614, \\ ignace@etsu.edu}

\author {\sc S.~Hubrig, }
\affil{
       European Southern Observatory, 
       Casilla 19001, 
       Santiago 19,
       Chile }

\author {\sc M.~Sch\"{o}ller}
\affil{
	European Southern Observatory, 
	Karl-Schwarzschild-Str.\ 2, 
	D-85748 Garching bei
	Muenchen, Germany
       }

\begin{abstract}

We report on low resolution ($R\approx 3000$) spectropolarimetry of the
A0 supergiant star HD~92207.  This star is well-known for significant
spectral variability.  The source was observed on seven different nights
spanning approximately 3 months in time.  With a rotation period of
approximately 1 year, our data covers approximately a quarter of the
star's rotational phase.  Variability in the continuum polarization
level is observed over this period of time.  The polarization across the
H$_\alpha$ line on any given night is typically different from the degree
and position angle of the polarization in the continuum.  Interestingly,
H$\beta$ is not in emission and does not show polarimetric variability.
We associate the changes at H$_\alpha$ as arising in the wind, which is in
accord with the observed changes in the profile shape and equivalent width
of H$_\alpha$ along with the polarimetric variability.  For the continuum
polarization, we explore a spiral shaped wind density enhancement in the
equatorial plane of the star, in keeping with the suggestion of Kaufer
\etal\ (1997).  Variable polarization signatures across H$_\alpha$
are too complex to be explained by this simple model and will require a
more intensive polarimetric follow-up study to interpret properly.

\end{abstract}


\section{INTRODUCTION}

Spectropolarimetry is a valuable tool for studying a number of important
properties of stars.  Circular polarization is one of the few direct
means of measuring stellar magnetic fields (e.g., Babcock 1958).  Linear
polarization, such as from Thomson scattering in circumstellar
envelopes of hot stars, has been important for ascertaining deviations
from sphericity in circumstellar media (e.g., the Be stars [Poeckert \&
Marlborough 1976; McLean \& Brown 1978; Wood, Bjorkman, \& Bjorkman 1997]
and supernovae [Wang \etal\ 1996; Leonard \etal\ 2005]).  A challenge
for this area is that net polarizations tend to be small so that large
telescopes are required to obtain sufficiently high quality data for
sources that are only moderately faint.  Fortunately, astronomers now
have access to several 8--10~m class or larger telescopes that are outfitted
with polarimeters.  With such instrumentation, studies of polarizations
across spectral lines -- combining good spectral resolution with large
telescope collecting areas -- have become increasingly popular and
important for understanding structure in circumstellar media.  Here we
present an analysis of variable linear polarization across the H$_\alpha$ line
and in the adjacent continuum
for the A0 supergiant star HD~92207 at seven epochs over 
a period of three months.

This particular star was chosen for several different reasons.
It has been studied spectroscopically by Kaufer \etal\ (1996) who
found substantial line profile and line equivalent width changes at
H$_\alpha$.  These variations appear to be related to the well-known
discrete absorption components (or ``DACs'') that are commonly seen in UV
lines of early-type stars (e.g., Massa \etal\ 1995; Kaper \etal\ 1996).
HD~92207 is not far removed from the Luminous Blue Variable stars and
O supergiants that show variable continuum polarizations (Lupie \&
Nordsieck 1987; Taylor \etal\ 1991; Harries, Howarth, \& Evans 2002;
Davies, Vink, \& Oudmaijer 2007).  Moreover, the relatively late spectral
type of HD~92207 within the early-type class indicates that H$_\alpha$
acts more nearly as a scattering line than a recombination line
(Puls \etal\ 1998),
thus making the star a prime candidate for exploring variable polarization
across a line dominated by wind emission.

A description of the observations is presented in the following
section.  An analysis of the data is provided in section~3.  A
discussion of the implications of the results is given in section~4.

\section{OBSERVATIONS AND DATA REDUCTION}

The observations of the supergiant HD\,92207 were obtained
in service mode from 2007 January to 2007 March at the European
Southern Observatory with FORS\,1 (FOcal Reducer low dispersion
Spectrograph), mounted on the 8-m Kueyen telescope of the VLT.  This
multi-mode instrument is equipped with polarization analyzing optics
comprising super-achromatic half-wave and quarter-wave phase retarder
plates, and a Wollaston prism with a beam divergence of 22\arcsec{}
in standard resolution mode. To perform linear polarization
measurements, a Wollaston prism and a half-wave retarder waveplate
were used.  The waveplate was rotated in 22.5$^\circ$ steps between
0$^\circ$ and 157.5$^\circ$, taking two 20\,s sub-exposures at each
of the eight positions.

The GRISM\,600R was used in the FIMS (FORS Instrumental Mask Simulator)
observing mode in the wavelength range 4672--6795\,\AA{} to cover the
hydrogen Balmer lines H$_\alpha$ and H$_\beta$.  The spectral resolution
of the FORS\,1 spectra taken with this setting and a 0.4\arcsec{} slit
was R$\sim$3000, corresponding to a velocity resolution of about
100 km~s$^{-1}$. 

The readout time was reduced to about 40\,s by windowing the CCD,
and the use of a non-standard readout mode (A,1$\times$1,low)
provided a broader dynamic range, allowing us to increase the S/N
of the individual spectra.  Due to the brightness of the target
(m$_V=5.5$) and excellent seeing conditions, a few exposures obtained
during the second and the last observing night were saturated.  For
these phases, only measurements at the retarder waveplate positions
between  0$^\circ$ and 67.5$^\circ$ are available.

\begin{figure}[t]
\centering{\epsfig{figure=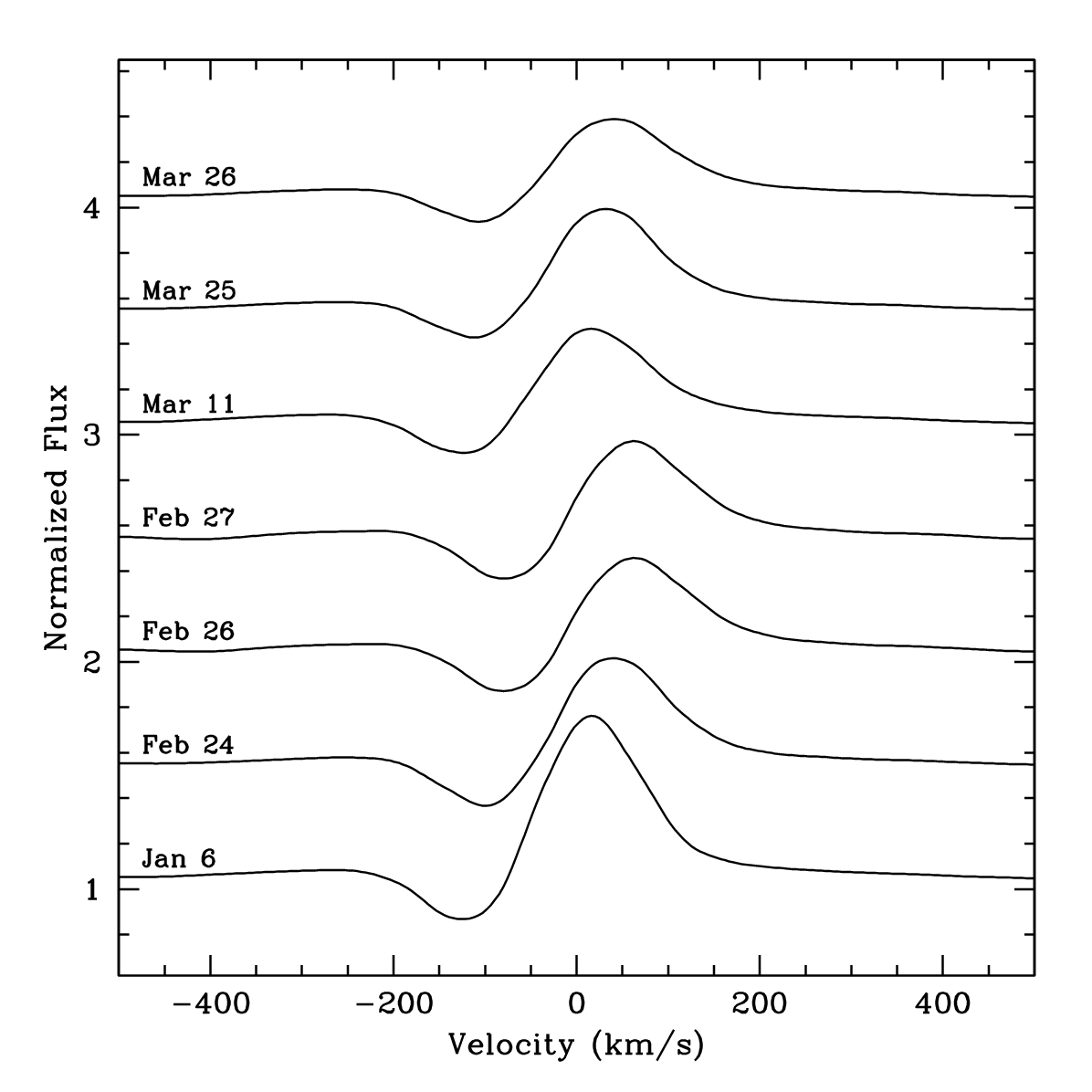, height=8.0cm}}
\caption{\small H$_\alpha$ spectra plotted as normalized to the local
continuum and shifted in the vertical
direction for clarity.  The epochs are indicated for
each line.  Note that the data have been smoothed.  Substantial variations
in profile shape are evident, with line equivalent widths
given in Tab.~\ref{tab1}.  The peak emission clearly shifts
between epochs.  
\label{fig1}}
\end{figure}

\begin{table*}
  \centering
  \caption[]{
Observing Journal
} 
\label{tab1}
  \centering
  \begin{tabular}{@{}ccccc@{}}
    \hline\hline
    Date & MJD & H$_\alpha$ EW & $Q_{\rm cont}^a$ & $U_{\rm cont}^a$  \\ 
         &     & ($\AA$) & (\%) & (\%) \\ \hline
    2007 Jan 06 & 54106.356& --4.02 & $-1.96 \pm 0.04$ & $-2.55 \pm 0.04$ \\  
    2007 Feb 24 &54155.189  & --3.16 & $-2.12 \pm 0.04$ & $-2.60 \pm 0.04$ \\
    2007 Feb 26 &54157.112  & --2.91 & $-2.14 \pm 0.04$ & $-2.64 \pm 0.04$ \\
    2007 Feb 27 &54158.117  & --2.60 & $-2.13 \pm 0.04$ & $-2.63 \pm 0.04$ \\
    2007 Mar 11 &54170.291  & --3.50 & $-1.90 \pm 0.04$ & $-2.51 \pm 0.04$ \\ 
    2007 Mar 25 &54184.162 & --3.47 & $-1.79 \pm 0.04$ & $-2.60 \pm 0.04$ \\ 
    2007 Mar 26 & 54185.151 & --2.96 & $-1.87 \pm 0.04$ & $-2.55 \pm 0.04$ \\
    \hline 
     \end{tabular}
\begin{center}
{\small $^a$ The continuum polarizations in the final
two columns are taken from the continuum
intervals of $6460-6470\AA$ and $6630-6650\AA$ adjacent to the
H$_\alpha$ line.}
\end{center}
\end{table*}

For all seven nights we calculated the values $P_Q = Q/I$ and $P_U =
U/I$, where $I$, $Q$, and $U$ are the Stokes parameters as defined
in Shurcliff (1962).  The total linear polarization is given by $P
=\sqrt{P_Q^2+P_U^2}$, and the polarization position angle is $\psi=
0.5\, \tan^{-1} (P_U/P_Q)$.  The associated errors for the determination
of $P_Q$, $P_U$, $P$, and $\psi$ are estimated from error propagation,
based on pure photon noise in the raw data (see Tab.~\ref{tab1}).

The spectropolarimetric calibration of FORS\,1 was checked with
the spectropolarimetric standard NGC\,2024-1.  On 2007 January 8,
300\,s exposures were taken at retarder waveplate angles 0$^\circ$,
22.5$^\circ$, 45$^\circ$, and 67.5$^\circ$.  These data were reduced in
the same manner as those obtained for our target HD\,92207.  We obtained
the values $P$ = 9.65$\pm$0.12\% and $\psi$ = 136.82$\pm$0.34$^\circ$,
which are in very good agreement with the previous measurements $P$ =
9.53$\pm$0.02\% and $\psi$ = 136.75$\pm$0.16$^\circ$ by Fossati \etal\
(2007), who used numerous spectropolarimetric observations of NGC\,2024-1
retrieved from the ESO archive.

\section{ANALYSIS}	\label{sec:analysis}

Here we describe the characteristics of the observed polarizations
and their implications for the source.  Based on previous studies of
H$_\alpha$, it is not surprising that even in low resolution spectra,
the P Cygni emission line of H$_\alpha$ of HD~92207 shows significant
variability, both in line shape and equivalent width.  Figure~\ref{fig1}
shows average line profile shapes for the seven spectra with dates
labeled.  The different spectra have been shifted vertically for
better display.  The equivalent widths (EWs) were evaluated between
$\pm 275$ km~s$^{-1}$, the terminal speed $v_\infty$ of the wind (see
Tab.~\ref{tab2}), and are listed in Table~\ref{tab1}.  Note that all of
the lines show a net emission.

\begin{figure*}[t]
\centering{\epsfig{figure=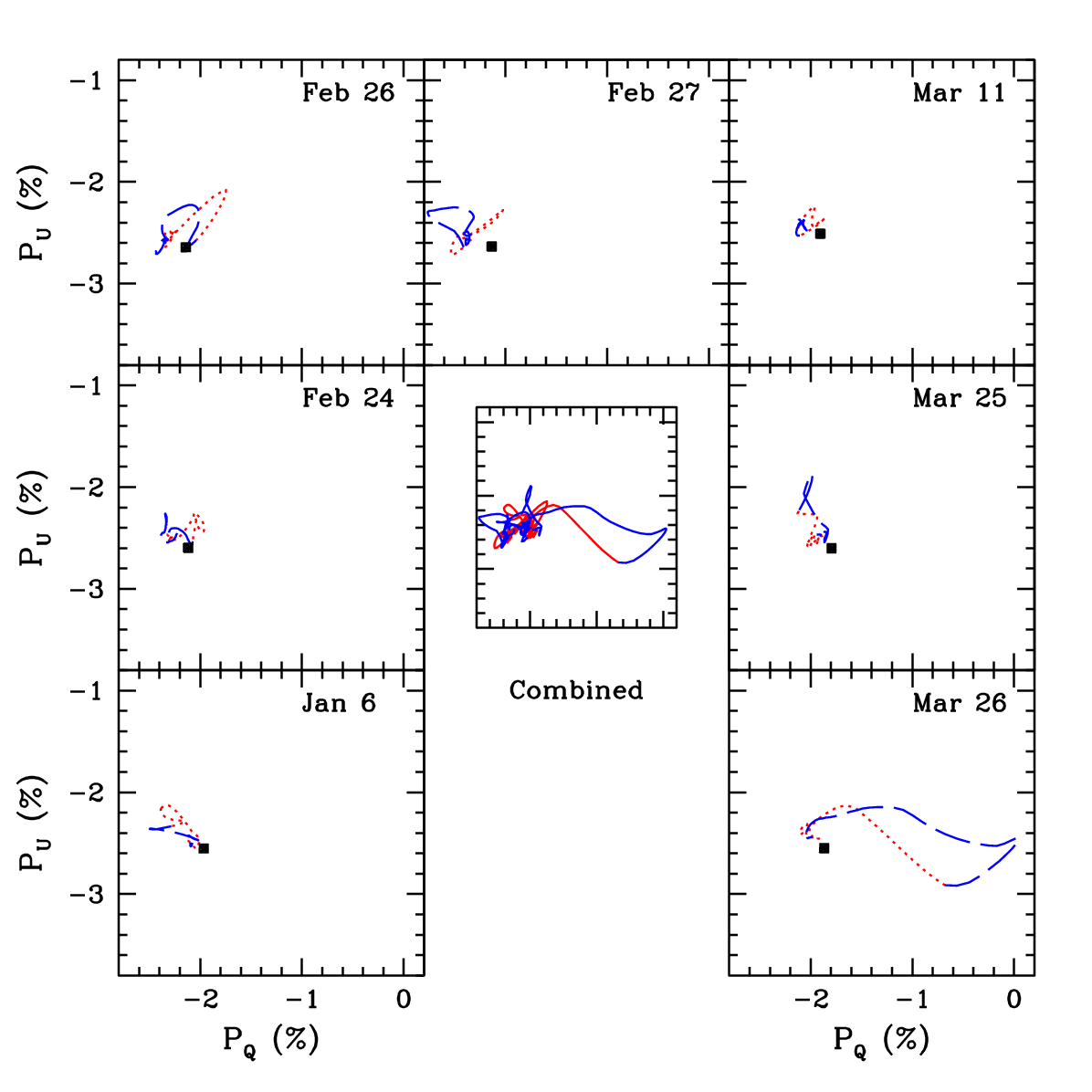, height=11.0cm}}
\caption{\small The variable polarization across the H$_\alpha$
line is displayed here in separate panels with dates given.  The center
panel overplots the various QU-loops.  The line types are dashed
for wavelengths blueward of line center, and dotted for those redward.
(The online version shows blue and red colors corresponding to the
velocity shift from line center.)
The black points are the nightly mean Q-U values of the nearby continuum. 
Notably, variations between the polarization in the line
and in the continuum is generally present, but to varying degrees,
and also at 
different position angles.
\label{fig2}}
\end{figure*}

Figure~\ref{fig2} shows the polarization of H$_\alpha$ and the neighboring
continuum in the form of Q-U diagrams for each night.  Each panel
displays the date, moving chronologically clockwise from lower left.
The black point in each panel signifies the mean continuum polarization
from relatively ``clean'' regions of the spectra that appear to lack
lines, including 6545--6557~\AA\ on the shortward side of H$_\alpha$
and 6569--6580~\AA\ on the longward side.  We determined a typical
standard deviation in the continuum polarization to be approximately
$\sigma \approx 0.04\%$ for any given night, which is approximately
the size of the plotted black squares.  Combining all seven nights,
the average polarizations are $\langle P_Q\rangle = - 1.99\%$ and
$\langle P_U\rangle= -2.58\%$; however, no attempt has been made to
correct the data for interstellar polarization, so these values cannot
be considered intrinsic to the star.  However, our analysis focuses on
variable polarization which is intrinsic to the star.

For wavelengths within the line, dashed and dotted line types distinguish
between blue and red shifts from line center (these appear as blue and red
colors in the online version of the figure).  Again, points considered to
be within the line are those between $\pm 275$ km~s$^{-1}$ of line center.

The central panel is an overplot of the line data for the other seven
panels, indicating the full range of polarimetric variations across
H$_\alpha$.  There is a change in the continuum polarization over the 3
month time span.  The overall amplitude of change is about 0.5\%, which
is over 10 times larger than the dispersion for any one of the averages.

Most remarkable are the observed variations in the polarization across
the H$_\alpha$ P Cygni line.  The average of the polarization within
the line is typically greater than the dispersion in the continuum knot
by a factor of 4 or more.  The two main exceptions are the nights of
March 11 that shows very little variation across the line and March 26
that shows the greatest variations, essentially a change in
polarization of about 2\%.  

It is notable that the variations across H$_\alpha$ imply different
polarization position angles on different nights.  Recall that $\tan\
(2\psi) = P_U/P_Q$.  Consequently, for a fixed geometry, $\psi$ will be
a constant, implying that $P_U \propto P_Q$ which amounts to variations
along a line in Q-U diagrams.  Although this seems approximately true for
the wavelength-dependent polarization across H$_\alpha$ on any given night
(February 26 being an exception), the extensions of the polarization from
the knot of continuum points appear to fall along different orientations
between nights.  Take for example the consecutive nights of March 25
and 26.  Polarimetric changes are in U for the former, but in Q for the
latter, suggesting a rotation in $\psi$ by $90^\circ$ after just one day.
Given that the size and rotation of the star suggest a rotation period
of approximately 1 year (see Tab.~\ref{tab2}), variations at the level of
1 day seems far too short to arise from a substantial change in geometry.

\section{DISCUSSION}

We have presented relatively medium resolution spectropolarimetric data
of the highly variable A supergiant HD~92207.  It has been suggested
by Kaufer \etal\ (1997) that the observed photometric and H$_\alpha$
line variations are the result of a corotating structure in the wind,
which they consider to be in the star's equatorial plane.  To explore
that possibility, we begin with a consideration of the variable
{\em continuum} polarization.  

We have developed a phenomenological model for Thomson scattering from
an enhanced scattering region in a spherical wind in the form of a
spiral pattern, similar in spirit to the work of Brown \etal\ (1994)
who sought to model DACs in emission lines.  Our objective is to obtain
a reference model for the global wind morphology that we can use for
interpreting the H$_\alpha$ polarization.

We consider a corotating stucture as a simple spiral that is top-bottom
symmetric about the plane of the star's rotational equator.  Hence, the
spiral structure has a guiding center that is always in the equatorial
plane.  This center obeys an equation of motion for the radial wind
velocity law and conservation of angular momentum in the rotating frame,
so it follows a ``streak line'' (e.g., see Ignace, Bjorkman, Cassinelli
1998).  We adopt a standard ``$\beta$-law'' for the radial flow:

\begin{equation}
v_{\rm r} = v_\infty\,\left(1-bu\right)^\beta,
\end{equation}

\noindent where $u=R_\ast/r$ and $b<1$ determines the radial speed
of the wind at its base, with $v_0 = v_\infty (1-b)^\beta$.

\begin{table}[t]
  \centering
  \caption[]{
Model Parameters
\label{tab2}
} 
  \centering
  \begin{tabular}{@{}cc@{}}
\hline\hline
Factor & Value \\ \hline
$R_\ast^a$  & $223 R_\odot$ \\ 
$v_{\rm rot}\sin i^a$  & 30 km s$^{-1}$ \\ 
$\dot{M}^b$ & $1.3\times 10^{-6}\, M_\odot$ yr$^{-1}$ \\ 
$v_\infty^a$  & 275 km s$^{-1}$ \\ 
$\beta^c$  & 1 \\ 
$b^c$  & 0.98 \\ 
$\eta^c$  & 25 \\ 
$\delta^c$  & $55^\circ$ \\ 
$\psi_0^c$  & $8^\circ$ \\ 
$i_0^c$  & $70^\circ$ \\ 
\hline
\end{tabular}
\begin{center}
{\small $^a$ Przybilla \etal\ (2006); $^b$ Kudritzki \etal\ (1996); $^c$ best fit
model parameters (see text).}
\end{center}
\end{table}

Velocity laws with $\beta=1$ and $\beta=2$ were considered, and
the case of $\beta=1$ produced a better match to the data.  In this
case the location of the guiding center is given analytically with
azimuth $\varphi$ by 

\begin{eqnarray}
\varphi(u) & = & \varphi_0(t) -\frac{\Omega\,R_\ast}{v_\infty}\,\left\{
	\frac{1-u}{u} +b\ln\left[\frac{w(u)}{uw_0}\right]\right. \nonumber \\
 &  & \left. -\frac{1}{b}\,\ln\left[\frac{w(u)}{w_0}\right]\right\},
\end{eqnarray}

\noindent where $w_0=v_0/v_\infty$ and $\Omega = 2\pi/P$.

The density for the spiral-shaped perturbed region is treated as an
{\em excess} of density above the otherwise spherically symmetric wind.
This density excess in the spiral pattern is taken to scale with the
spherical wind density in our ``toy'' model, thus we conveniently
parametrize the excess by a constant factor $\eta = n_{\rm excess}/
n_{\rm sph}$, for $n_{\rm sph}$ the spherical wind density.  In addition
to the solution for the guiding center, we also need the cross-section
of the spiral.  The cross-section (i.e., the intersection of the spiral
pattern with a spherical shell) is treated as circular.  This spherical
``cap'' is axisymmetric, and so assuming the electron scattering is
optically thin, the polarization from any given slice of the spiral is
given by Brown \& McLean (1978), along with the finite star depolarization
correction factor of Cassinelli, Nordsieck, \& Murison (1987).  Summing up
contributions from all the caps yields the polarization from the structure
as a function of rotational phase and viewing inclination for the rotation
axis of the star.

Note that our model accounts for occultation of scattered light by
the intervening star, but only in an approximate way.  We consider a
slice as entirely occulted if its guiding center lies behind the star,
and unocculted otherwise.

\begin{figure}[t]
\centering{\epsfig{figure=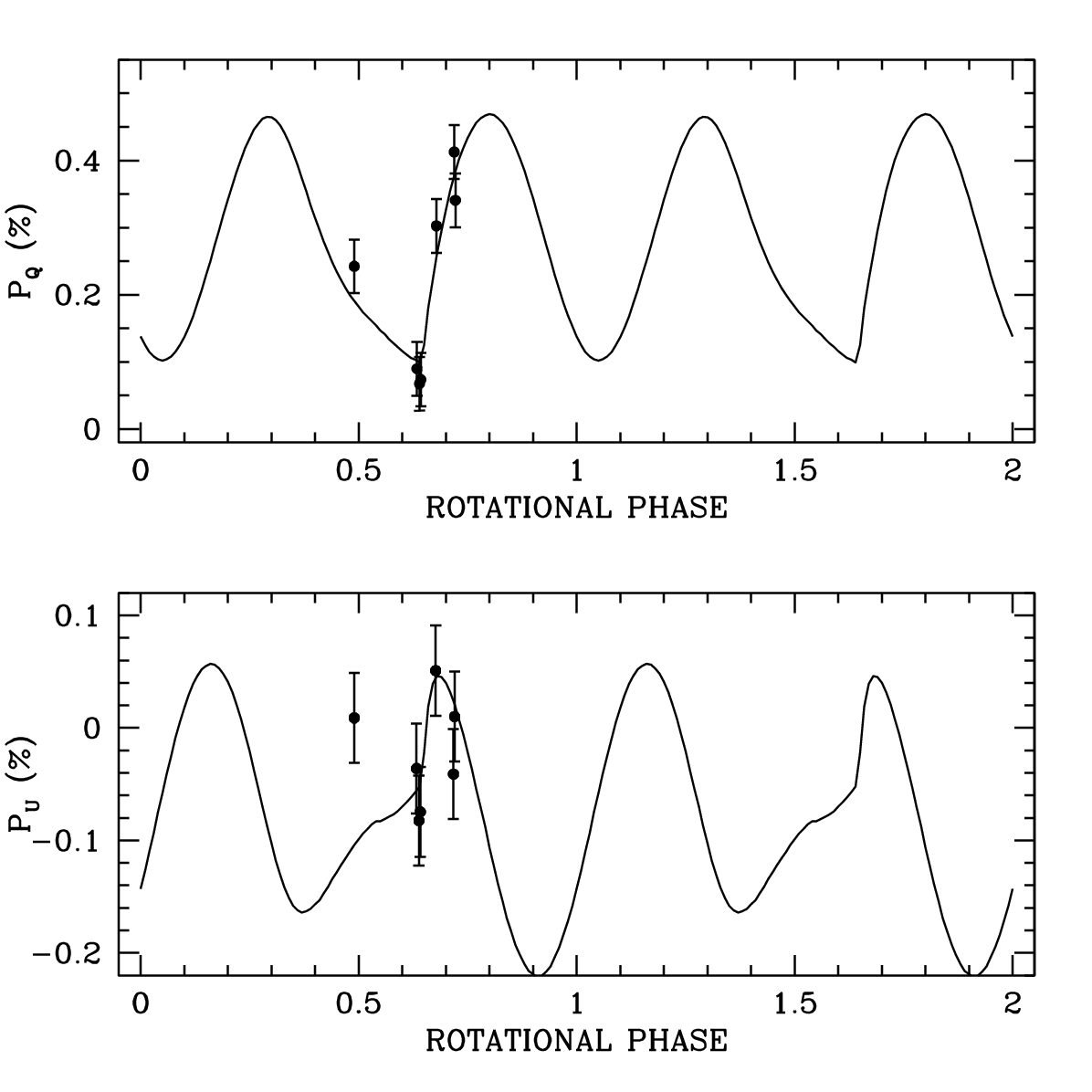, height=8.0cm}}
\caption{\small Model $P_Q$ (top) and $P_U$ (bottom) light curves plotted
for two rotational cycles based on the model described in the
text with parameters listed in Tab.~\ref{tab2}.  Observed
data are plotted as points with error bars.
The data points have been shifted horizontally and
vertically to provide the best fits.
\label{fig3}}
\end{figure}

The principal model parameters are the density excess $\eta$,
the half-opening angle of the spiral $\delta$, an orientation angle
between the observer $Q-U$ axes and those of the star system $\psi_0$,
and finally the inclination of the rotation axis of the star $i_0$.
Using a reduced chi-square evaluation for a grid of model polarization
light curves, Table~\ref{tab2} lists the model parameters that provide
the best fit to the observed {\em continuum} data.  The star and wind
properties of HD~92207 are taken from Przybilla \etal\ (2006), except
that the mass-loss rate is taken from Kudritzki \etal\ (1999) and does
not account for clumping.  The most reasonable match to the observed
continuum polarizations in the neighborhood of H$_\alpha$ is shown in
Figure~\ref{fig3}.  The upper panel is for $P_Q$ and the lower one for
$P_U$, displayed as percentage polarizations.  The rotational phases
depend on the star's rotation period.  Given the radius and minimum
rotation speed from Table~\ref{tab2}, the maximum period is $P_{\rm max}
\approx 376$\,d.  The true period is $P_{\rm rot} = P_{\rm max}/\sin i_0$,
where $i_0$ is constrained from our model fitting.  As the ephemeris
is not known, we assigned the rotational phase ``0'' to the date of
our first observation, and the phases appearing in Figure~\ref{fig3}
represent values for our best fit model at $i_0 = 70^\circ$.

For the model fitting, there are seven free parameters: the four listed
above plus three offsets -- one for rotational phase, a vertical offset in
$Q$, and an independent one in $U$.  With fourteen data points in total,
the reduced chi-square for our best simultaneous fit to the $P_Q$ and
$P_U$ lightcurves is 1.6.  That value is primarily a reflection of one
discrepant data point, the first one in the observed $P_U$ ligthcurve.
Our model is inherently smooth, whereas the line variability indicates
the presence of variable wind structure.  For the continuum polarization,
this wind structure can act as a source of ``noise''.  Better temporal
and phase coverage is needed to interpret the polarimetric
lightcurve more fully.

The non-zero variation of $P_U$ indicates that $i_0$ is not edge-on to
the system, since no $\Delta P_U$ could result for such a perspective
when the circumstellar scattering environment is top-bottom symmetric.
However, $\Delta P_U$ is much less than $\Delta P_Q$, suggestive that
$i_0$ is not close to pole-on.  Since the interstellar polarization is
not known, there is freedom to shift the observed points with respect to
the model in both phase (i.e., $t_0$) and vertical offset to best match a
model curve.  The vertical offsets for $P_Q$ and $P_U$ are independent,
but the phase shift must be the same for both.  There is also freedom
to rotate the model system relative to the observer $Q$-$U$ system,
but we found the angle to be small, with $\psi_0 = 5^\circ$.

\begin{figure*}[t]
\centering{\epsfig{figure=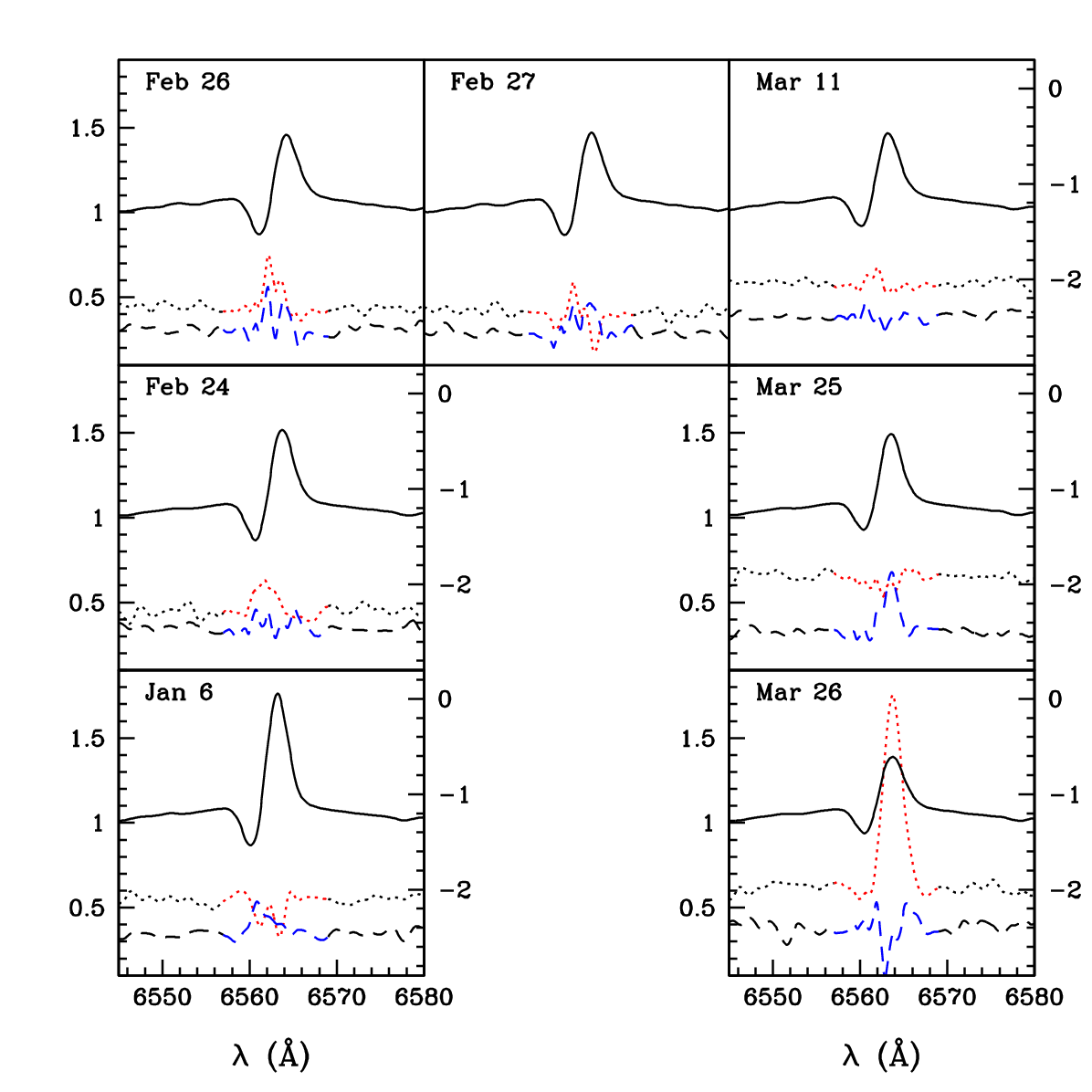, height=11.0cm}}
\caption{\small An overplot of the P Cygni lines against the
$P_Q$ (dotted curve, or red for the online version)
and $P_U$ (dashed curve, or blue for the online version)
spectra using the format of Fig.~\ref{fig2}
to highlight the velocity locations of features.
The H$_\alpha$ line
is normalized to the local continuum, and plotted as the solid curve
using the scale on leftside axes; polarizations are shown as percent
using the scale on the rightside axes.
\label{fig4}}
\end{figure*}

The match between the model and the continuum data appears to be best
around an inclination of $i_0=70^\circ$ with a half-opening angle of
$\delta=55^\circ$.  A $55^\circ$ opening angle corresponds to the Van
Vleck angle and yields the maximum polarization when other parameters are
held fixed, thus allowing us to minimize the value of the excess density.
Because the spherical wind is relatively thin in electron scattering
($\tau_{\rm e} \approx 0.014$ for solar abundances with ionized H and
He neutral), a fairly large value of $\eta=25$ is needed to match the
observed $\Delta P_Q$ and $\Delta P_U$.  Such a large excess would need to
be justified through physical modeling of the stellar wind and atmosphere.

Overall, the match is reasonably good except for the first data point
in the lower panel for $P_U$ which we cannot reconcile to the model
after exploring a range of opening angles, viewing inclinations, and
$\eta$ values.  Based on the work of Davies \etal\ (2005) for
luminous blue variables, wind inhomogeneities are likely contributing a
random contribution of unknown amplitude to the observed polarizations
from HD~92207, which makes the interpretation of any globally coherent
variation with phase challenging for such sparse sampling.  Even though
our spiral model is simplistic, it does capture the essential points
inherent to any corotating structure in terms of the generic trends of $P_Q$
and $P_U$ with phase, viewing inclination, and density.  It is certainly
clear that much better sampling of the polarimetric light curve in the
continuum is needed to critically assess the existence of a corotating
region in the wind.

Turning to the H$_\alpha$ line, the P Cygni morphology indicates that
the line is dominated by the stellar wind.  In contrast, the H$_\beta$
line is in absorption and shows no variable polarization.  At the
spectral type of HD~97702, the $n=2$ level of hydrogen can act as the
effective ground state (Puls \etal\ 1998) explaining why it shows such a
strong P~Cygni shape.  It also means that line scattering polarization
(e.g., Hamilton 1947) may influence the observed variable polarization
in H$_\alpha$.  For example, such effects were explored by Jeffery (1987,
1989) in the case of SN1987A.

Figure~\ref{fig4} displays the polarized $P_Q$ and $P_U$ spectra in
velocity along with the P Cygni profiles to emphasize where polarization
changes occur within the line in relation to the absorption trough and
emission peak.  The vertical scale for the measured percent
polarizations are 
shown on the right axes, whereas the scale
for the continuum normalized P Cygni lines is displayed on the left axes.
For the first 5 panels, it is typically the case that either $P_Q$, $P_U$,
or both vary most around the line core over an interval that is about a
third of the wind terminal speed.  One clear exception is for March 26
that shows the strongest change in polarization across the line in $Q$
and aligns quite well with the redshifted emission peak, although it
appears that $P_U$ actually increases in absolute value.  The observed
polarizations are negative, so zero polarization is toward the top of
each panel.  Curiously, there is a depolarization at the same location the
night prior, except in $U$ instead of $Q$.  The overall depolarizations
across the emission peak in the line relative to the continuum would
seem to be a classic ``line effect''.  Polarization is normalized at
each wavelength to the total emission.  If the line formation leads
to unpolarized radiation that is largely unscattered by free electrons
before emerging from the wind, then the additional line emission tends
to reduce the polarization at those wavelengths relative to values
outside the line.  In fact, this can be an excellent way of placing
an upper limit to the interstellar polarization to the star.  However,
the H$_\alpha$ line of HD~92207 is perplexing because its behavior is
not consistent.  Moreover, night-to-night variations in the H$_\alpha$
polarization are hard to understand in terms of the spiral structure
that we have considered for the continuum polarization given that the
star's rotation period is of order a year.

It is worth pointing out that Harries (2000) modeled wavelength-dependent
polarization across H$_\alpha$ for the O supergiant $\zeta$~Pup.
His models did not include contributions to the polarization from line
scattering; rather, he obtained changes of polarization across the line
owing to the influence of line opacity for the polarization produced
by Thomson scattering.  

The application of Harries' models to HD~92207 is unclear.  There is the
likelihood of a significant line scattering contribution to H$_\alpha$
because of HD~92207's much later spectral type as compared to $\zeta$~Pup.
Other effects such as dynamic changes in the wind or variable stellar
illumination (c.f., Al-Malki \etal\ 1999) will be needed to interpret
the polarization across H$_\alpha$.  For example, it would be useful
to combine the methods of Li \etal\ 2000 and Davies \etal\ 2007 for
polarimetric variability from wind clumping and electron scattering
with the resonance scattering polarization effects explored by Ignace,
Nordsieck, \& Cassinelli (2004) for stationary winds.  With limited
coverage of the rotational phase and only modest spectral resolution,
the existing dataset is too poor to undertake a detailed calculation
of the H$_\alpha$ line to interpret the observed polarizations.  With
future data, such an effort would be worthwhile because the continuum
polarization constrains the global wind morphology whereas the H$_\alpha$
line is sensitive to vector velocity flow.

\section*{Acknowledgements}

We are grateful to an anonymous referee for a number of helpful suggestions.
RI thanks Gary Henson for numerous discussions about the polarization
changes across lines.  This work was partially supported by a grant from
the National Science Foundation (AST-0807664).  We want to acknowledge
the ESO program 078.D-0330(A) that made the observations possible.

\end{document}